\documentclass[noshowpacs,prl,aps,superscriptaddress,floatfix, twocolumn,footinbib]{revtex4-2} 
\usepackage{cuted}
\usepackage[utf8]{inputenc}
\usepackage{lineno,mathtools,url}
\usepackage{amsmath}
\usepackage{soul}
\usepackage{amssymb,mathrsfs}
\usepackage{physics}

\usepackage{graphicx}    \usepackage{cuted} 
\usepackage{hyperref}

\usepackage{comment}
\usepackage{dsfont}
\usepackage[dvipsnames]{xcolor}
\hypersetup{
     colorlinks=true,
     linkcolor=blue,
     filecolor=blue,
     citecolor = orange,      
   }

\newcommand{\px}{\partial_x}
\newcommand{\pt}{\partial_t}

\usepackage{mathbbol} 
\usepackage[normalem]{ulem}

\newcommand{\lt}{\left(}
\newcommand{\rt}{\right)}

\allowdisplaybreaks
\begin{document}

\title{
Anomalous Doppler effect in two-component Bose-Einstein condensates}

\author{Tomasz Zawi\'slak}
\affiliation{Pitaevskii BEC Center, CNR-INO and Dipartimento di Fisica, Universit\`a di Trento, Via Sommarive 14, 38123 Povo, Trento, Italy}
\author{Sandro Stringari}
\affiliation{Pitaevskii BEC Center, CNR-INO and Dipartimento di Fisica, Universit\`a di Trento, Via Sommarive 14, 38123 Povo, Trento, Italy}
\author{Alessio Recati}
\affiliation{Pitaevskii BEC Center, CNR-INO and Dipartimento di Fisica, Universit\`a di Trento, Via Sommarive 14, 38123 Povo, Trento, Italy}
\affiliation{INFN-TIFPA, Trento Institute for Fundamental Physics and Applications, I-38123 Trento, Italy}

\begin{abstract}
We show that two-component Bose-Einstein condensed mixtures, in presence of a persistent current, exhibit a non-trivial Doppler shift of the sound velocities. The peculiarity is due
to the interspecies interaction and the possibility of generating a counter-flow persistent
current. Analytical predictions are derived using superfluid hydrodynamics. While the existence of anomalous Doppler shifts at finite
temperature has been discussed a long time ago in the case of superfluid Helium-4, an experimental verification
of the effect is still missing. For this reason, we also propose a protocol for the measurement of
the Doppler shifts, based on the density-density response function. The dynamical protocol is simulated by means of coupled Gross-Pitaevskii equations.

\end{abstract}

\maketitle

\section{Introduction}
In an ordinary fluid, moving as a whole   at  velocity $v_f$, the speed of sound, $c^0$, is modified according to the Galilean transformation which puts the fluid in a reference frame where it is at rest. Simplifying the discussion to one dimensional configurations the speed of sound in the lab frame reads $c^\pm = c^0 \pm v_f$, where the $\pm$ refers to phonons co- or counter-propagating with respect to the fluid flow. The effect is referred to as the kinematic Doppler shift. 

Almost 70 years ago, Khalatnikov pointed out that, for the superfluid state of He-4,  two-fluid hydrodynamics predicts an anomalous, non-kinematic Doppler effect, due to the relative motion between the normal and the superfluid components \cite{Khalatnikov56}. A more detailed analysis of the different sound modes was carried out in the 90s by Nepomnyashchy and Revzen \cite{NepoHeII,NepoHe4thSound,NepoHe34}, but up to now, the effect has eluded direct experimental evidence. Recently, we have shown that the anomalous Doppler effect persists even at zero temperature for density-modulated superfluids, including supersolids \cite{dopplerSupersolid}, where the crystalline structure acts as a normal component (see, e.g., the recent work \cite{saslowInertia}).  In \cite{dopplerSupersolid}, we also proposed a protocol to measure the anomalous Doppler effect in cold gases platforms, for which a microscopic description, by means of Gross-Pitaevskii equations (GPEs), is possible.

In the present paper, we show that a binary mixture of weakly interacting Bose-Einstein condensates also exhibits a non-kinematic Doppler effect, as a consequence of the interspecies interaction which couples the motion of the two fluids. 
Bose-Einstein condensed mixtures, routinely realized in cold gases laboratories, are likely the best candidate to experimentally detect the anomalous Doppler shift of sound modes. In contrast to previously studied systems, a binary mixture of Bose-Einstein condensates does not involve a normal part, as it consists of two superfluids. This difference is essential for the stability of the relative current between the two fluids, a crucial requirement for the occurrence of  a non-kinematic Doppler effect. Since the superfluid velocity is set by the boundary conditions of the phase of the order parameter, a counterflow of two superfluids is expected to be more robust against noise of various sources, inevitably present in every experiment.

The paper is organized as follows. An analytical expression for the Doppler shift of the sound speeds is obtained using collisionless superfluid hydrodynamics. Subsequently, a protocol to properly measure the Doppler shift of the two sound modes is presented by using the time-dependent Gross-Pitaevskii equations. In particular, we develop a selective approach that permits the independent measurement of the Doppler shift of each sound by analyzing the precession of the total density. We finally discuss how the presence of current-current interspecies interactions, responsible for the so-called Andreev-Bashkin drag \cite{ABeffect}, affects the occurrence of the anomalous Doppler effect.

\section{Hydrodynamic model of superfluid mixtures}\label{sec:hd}
In this work, we consider a mixture of two atomic  Bose-Einstein condensates  occupying  two different hyperfine levels $|i\rangle$, $i=1,2$, of atoms with mass $m$. In the weakly interacting case, the atoms interact only via contact interaction potentials, which are characterized by the intraspecies coupling strengths,  $g_{11}$, $g_{22}$ and by the interspecies coupling strength $g_{12}$. They are related to the $s$-wave scattering lengths $a_{ij}$ via the expression $g_{ij}=4\pi\hbar^2 a_{ij}/m$.  In order for the mixture to be stable, the intraspecies interaction $g_{ii}$ must be positive and,  without loss of generality, we also assume $g_{12}>0$ in the rest of our work. 

In the following, we use a description of the mixture in terms of density and polarization or spin density. Given the single-component densities $n_{i}$, we define the total density $n=n_1+n_2$ and the spin density  $s_z=n_1-n_2$.  For a generic homogeneous Bose mixture, the equation of state can be written as \begin{equation}
\label{eq:energy}
\epsilon={1\over 2}g_{d} n^2+{1\over 2} g_{s} s_z^2+ {1\over 2}g_{ds} n s_z,
\end{equation}
where we have defined the following combinations of the coupling constants:
$g_d=(g+g_{12})/2$ and $g_s=(g-g_{12})/2$ with $g=(g_{11}+g_{22})/2$ as well as $g_{ds}=(g_{11}-g_{22})/2$.  The stability of the system requires the energy to be a convex function of $n$ and $s_z$, which gives rise to the further condition
\begin{equation}\label{eq:def_delta}
   \Delta= 4 g_d g_s-g_{ds}^2\ge 0.
\end{equation}
If the inequality Eq. (\ref{eq:def_delta}) is violated, the gas is unstable towards phase separation \footnote{The stability in terms of the intra- and interspecies interactions reads: $g_1 g_2-g_{12}^2\ge 0$. For $g_{12}<0$ the instability would lead to the collapse of the mixture.}. 
%In the present work we require the system to be miscible, hereafter assuming that the stability condition is always met. 

In the absence of permanent currents, the hydrodynamic theory of superfluids predicts the propagation of two sounds where the two fluids oscillate in phase (density-like sound) and out of phase (spin-like sound). The corresponding sound velocities (see, e.g., \cite{BecBook2016,DuineHD2BEC}) can be written as \footnote{In the single-component notation, the two speeds of sound have the familiar form: $c_{d/s}^0 = \frac{1}{\sqrt{2m}} \left[g_{11}n_1 + g_{22}n_2 \pm \sqrt{(g_{11}n_1 - g_{22}n_2)^2 + 4n_1n_2g_{12}^2}\right]^{\frac{1}{2}} $} 
\begin{equation}\label{eq:c0s}
    c_{d/s}^0=\left[\frac{g n +g_{ds} s_z\pm\sqrt{(g n +g_{ds} s_z)^2-\Delta (n^2-s_z^2)}}{2 m}\right]^{1/2}
\end{equation} 
and have been found in good agreement with experiments \cite{Gabriele2BECsound,Shin2BECsound,piekarski2BEC}. The discretized spin mode oscillations have also been  investigated in \cite{spindipole2016}.

The nature of the two sounds is rather complex as they are strongly hybridized.
In the limit of a non-interacting mixture, when $g_{12}=0$, the equation (\ref{eq:c0s}) reduces to $\sqrt{g_{ii}n_{i}/m}$
and the two modes coincide with  two single-component density oscillations. On the other hand, when the gas exhibits  $\mathbb{Z}_2$ symmetry ($g_{ds}=s_z=0$), the two sound modes coincide with pure total density and spin density oscillations. For the sake of notation, we use the $d$ ($s$) subscript to refer to the sound mode (\ref{eq:c0s}) with the $+$ ($-$) sign and call it \textit{density-like} (\textit{spin-like}) mode \footnote{If  $g_{12}$ were negative $\pm$ in the Eq. (\ref{eq:c0s}) has to be replaced by $\mp$ .}.

In the presence of permanent currents, the two speeds of sound will be modified. We determine such a shift by using the zero-temperature superfluid HD equations in one dimension:
\begin{subequations}\label{eq:hd_mixtures} 
\begin{align}
    \pt n+\partial_x j     &=0  \label{eq:hd_n}\\
    \pt s_z+\partial_x j_z &=0  \label{eq:hd_sz}\\
    m\pt v_T+\partial_x\left( \frac{mv_T^2}{2}+\frac{m w^2}{2}+\mu\right)&=0 \label{eq:hd_v} \\
    m\pt w+\partial_x\left( m wv_T+h\right)&=0 \label{eq:hd_w}
\end{align}
\end{subequations}
where we have introduced the chemical potential $\mu=\partial\epsilon/\partial n=g_d n+g_{ds} s_z/2$, the polarizing field $h=\partial\epsilon/\partial s_z=g_s s_z+g_{ds} n/2$, the average velocity $v_T=(v_1+v_2)/2$ and $w=(v_1-v_2)/2$, with $v_i$ the superfluid velocity of the $i$-th component. The density and the spin density currents are given by:
\begin{subequations}\label{eq:hd_js} 
\begin{align}
    j=&\ n v_T + s_z w \label{eq:hd_j} \\
    j_z=&\ s_z v_T + n w  \label{eq:hd_jz}  
\end{align}
\end{subequations}
and correspond respectively to the one-half of the sum and difference of the two single-component currents $j_i = n_i v_i$ \footnote{Notice that $j_z$ is the $z$-component of the Noether SU(2) current 
\begin{equation}
\mathbf{j}_s=v\mathbf{s}+\frac{\hbar}{2m}\left(\frac{\mathbf{s}}{n}\times\partial_x \mathbf{s}\right),
\end{equation}
where the first term corresponds to the spin advection, while the second is the so-called quantum torque. In the case of a SU(2) symmetric atomic mixtures, i.e., $g_s=g_{ds}=0$, the full spin would be conserved, and spin HD reduces to Noether conservation law: $\pt \mathbf{s}+\partial_x\mathbf{j}_s=0 $.}.
We write the velocity fields as $ v_T(x,t)=v_T^0+\delta v_T(x,t) $  and $w(x,t)= w^0+\delta w(x,t)$, respectively, where the constant velocities $v_T^0$ and  $w^0$  correspond to the permanent currents. 
The sound modes are derived using a linearization procedure, where one considers small deviations from the equilibrium values of the variables entering the equations (\ref{eq:hd_mixtures}). We identify the set of four independent variables $\xi\in\{n,s_z,v_T,w\}$ and express them as the sum of their equilibrium values $\xi^0$ and small variations, slowly changing in space and time $\delta\xi(x,t) = \delta\xi \exp( i(qx-\omega t ))$. In the $q\rightarrow 0$ limit, the dispersion law is linear, and its slope $c = \omega/q$ defines the speed of sound. The linearized HD equations (\ref{eq:hd_mixtures}) are neatly represented in matrix form:
\begin{equation}\label{eq:matrix}
    \begin{bmatrix}
        v_T^0-c & w^0& n^0& s_z^0\\
        w^0 & v_T^0-c& s_z^0& n^0\\
        g_d & g_{ds}/2 & m(v_T^0-c)& mw^0\\
        g_{ds}/2 & g_s& mw^0& m(v_T^0-c) \\
    \end{bmatrix}
    \begin{bmatrix}
        \delta n\\
        \delta s_z \\
        \delta v_T \\
        \delta w 
    \end{bmatrix}
    =
    \begin{bmatrix}
        0\\
        0 \\
        0 \\
        0 
    \end{bmatrix},
\end{equation}
from which one finds four solutions \footnote{We determined the four solutions as the roots of the fourth-order polynomial, resulting from the determinant of Eq. (\ref{eq:matrix}).  }, corresponding to the two Doppler shifted pairs of modes. We are interested in the linear shift of the speeds with respect to their value in a fluid at rest, i.e., Eq. (\ref{eq:c0s}). Assuming $v_T^0,\;w^0\ll c_{d/s}^0$, we find the following results for the Doppler-shifted sound velocities
\begin{equation}
\label{eq:doppler}
    c_{d/s}^{\pm}=c_{d/s}^0 \pm \left(v^0_T+ \delta_{d/s}w^0\right),
\end{equation}
where the +(-) sign corresponds to the sound co-propagating (counter-propagating) with the current, and
\begin{equation}
    \delta_{d}=-\delta_s=\frac{g_{ds} +g m_z}{\sqrt{(g +g_{ds} m_z)^2 - \Delta(1-m_z^2)}},
    \label{eq:hd_delta}
\end{equation}
with  $m_z = s_z^0/n^0$ the spin polarization. The hydrodynamic theory reveals the highly non-trivial nature of the Doppler effect in a superfluid Bose mixture, which depends on all the thermodynamic parameters of the gas. As anticipated, the Doppler shift becomes trivial once the interspecies coupling constant  $g_{12}$ vanishes. In such a limit Eq. (\ref{eq:hd_delta}) reduces to $\delta_{d/s}= \pm 1$ and  Eq. (\ref{eq:doppler}) simply coincides with the kinematic Doppler shift for each component, i.e., $v^0_T+ \delta_{d/s}w^0=v^0_{1/2}$. The same result is trivially found in the single-component limit, i.e., $m_z = \pm 1$.

Aside from interspecies interaction, non-trivial effects in the Doppler effect require the presence of a relative motion between the two components of the mixture ($w^0 \neq 0)$. In a classical fluid, collisional effects prevent the formation of any stable mixture with two different velocity fields. Instead, in a superfluid, a state with relative velocity $w^0$ can be obtained by imposing different conditions on the phases of the two order parameters, satisfying the appropriate boundary conditions. 

\begin{figure*}[!ht]
    \centering
    \includegraphics[width=0.99\linewidth]{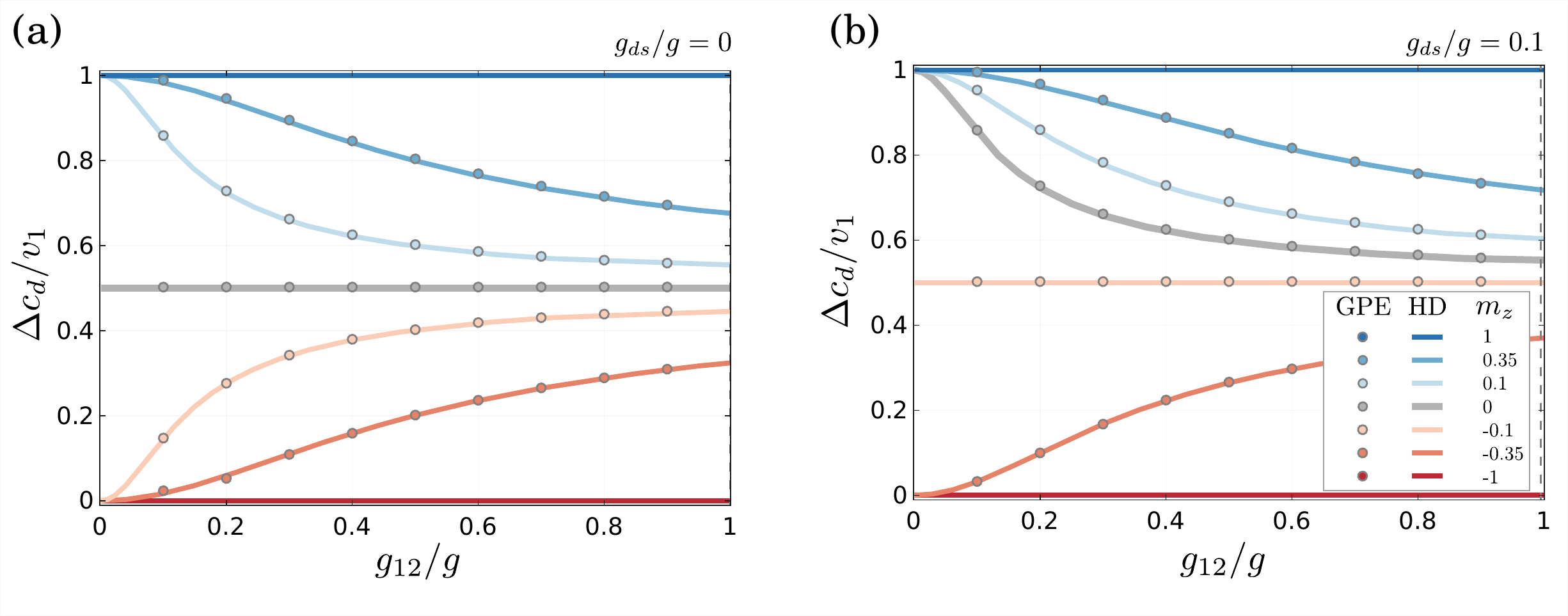}
    \caption{The Doppler shift of the density-like sound relative to the velocity of the first component as a function of interspecies interaction coupling $g_{12}$ for $v_2=0$ and \textit{(a)} $g_{ds}=0$ and \textit{(b)} $g_{ds}/g=0.1$ . Each solid line corresponds to a HD prediction for different values of polarization $m_z$. The symbols represent results of the GPE simulation, whose details are contained in the next section. The $g_{12}$ and $m_z$ dependence of the Doppler shift of the spin-like sound mode ($\Delta c_s$) is the same as of the density-like mode, but mirrored with respect to the $y=0.5$ line. The vertical dashed line marks the stability condition (\ref{eq:def_delta}).  }
    \label{fig:fig1}
\end{figure*}

Without loss of generality, in the following we assume that only the first superfluid component has a finite velocity $v_1$, while the second remains at rest. With this choice, the Doppler shifts of the two sounds read
\begin{equation}\label{eq:gpe_doppler}
    \Delta c_{d/s} = \frac{v_1}{2}\left( 1 + \delta_{d/s} \right).
\end{equation}
Moreover, since $\delta_{d/s}$ differs only in sign between the two modes, we focus only on the density-like one. In Fig.~(\ref{fig:fig1}a). we show the Doppler shift as a function of $g_{12}/g$ for $g_{ds}=0$ and different polarizations $m_z$.

The Doppler shift of the density-like sound is, as expected, always smaller than $v_1$ and reaches the value $\delta_{d/s}=\pm m_z$ by approaching the SU(2) symmetry point as $g_{12}\rightarrow g$.

It is intriguing to note that for every set of interaction constants, there exists a special value, $m_z = -g_{ds}/g$, for which the numerator of Eq.~(\ref{eq:hd_delta}) vanishes. Consequently, the Doppler shift of both sounds is given by $v_T^0 = v_1/2$ and is independent of the value of the coupling $g_{12}$. An example of this unusual dependence is shown in Fig.~(\ref{fig:fig1}b), where $g_{ds}/g = 0.1$. For the $\mathbb{Z}_2$ symmetric gas, such an effect is observed if $g_{ds}=0$ (see Fig.~(\ref{fig:fig1}a)). This result is true as long as the relative velocity is small enough, i.e., \begin{equation}\label{eq:condition}
m w^2\ll g_{12}n^0\sqrt{1-m_z^2},
\end{equation} which shows that for either $g_{12}\rightarrow 0$ or $m_z\rightarrow 1$ the peculiar results of having an equal Doppler shift for both sounds cannot hold. In fact, in the limit of small coupling constant $g_{12}$, the two sounds continuously change the character from density- and spin-like oscillations to single-component modes at $c_d^0-c_s^0 \approx v_T^0$. For $m_z=-g_{ds}/g$, the difference between the two speeds of sound at rest is comparable to the velocity $v_T^0$ when $g_{12}/g \approx \sqrt{2m/(gn^0)}v^0_T$. For the parameters used in Fig.~(\ref{fig:fig1}) this evaluates to $g_{12}/g\approx 0.005$, satisfying the condition (\ref{eq:condition}).

As mentioned before, the Doppler shift of the spin-like sound differs from the one of the density-like only in sign of the $\delta$ term (\ref{eq:def_delta}). This means that the entrainment of the spin-like sound has the same dependence on $g_{12}$ as the ones depicted in Fig.~(\ref{fig:fig1}a) and Fig.~(\ref{fig:fig1}b), but reflected with respect to the $y=0.5$ axis. 

% Another case showing no such adependence is the $\mathbb{Z}_2$ symmetric gas, where there is an equal number of particles of both kinds. Indeed, for $g_{ds} = 0$ and $s_z = 0$ the $\delta_{d/s}$ (which carries all the thermodynamic quantities entering the Doppler shift) reduces to $0$. In this special case, where the two sounds $c_d$ and $c_s$ become purely density and spin oscillations, in the presence of a finite velocity field $v_1$ in only one component, the two speeds of sound are shifted by the same value $v_T = v_1/2$. Then, the Doppler effect is effectively associated with a simple Galilean transformation to the reference frame moving with velocity $v_1/2$.

\section{Dynamic protocol to measure the Doppler effect}\label{sec:gpe}

In the following, we propose an experimental protocol to excite the sound modes and measure the Doppler shifts of both sounds. For this purpose, we employ two coupled Gross-Pitaevskii equations, which are equivalent to the hydrodynamic description (\ref{eq:hd_mixtures}) if one neglects the quantum pressure \cite{stringari96}. The GP equations are conveniently represented in the matrix form
\begin{eqnarray}\label{eq:2c_gpe}
i \hbar\frac{\partial}{\partial t}
    \begin{pmatrix}
    \psi_1 \\ \psi_2
    \end{pmatrix}
    =
    \begin{pmatrix}
    h_1 & 0\\ 0 & h_2
    \end{pmatrix}
    \begin{pmatrix}
    \psi_1 \\ \psi_2
    \end{pmatrix},
\end{eqnarray}
where $\psi_i$ is the order parameter of the $i$-th component, $h_i = -\frac{\hbar^2}{2m}\nabla^2 + g_{ii}n_i + g_{12}n_{3-i}$ is the corresponding single-component Hamiltonian, with $n_i(x,t)=|\psi_i(x,t)|^2$ the densities of the two components. We investigate solutions of Eq.~(\ref{eq:2c_gpe}) in one dimension, where periodic boundary conditions allow for the existence of permanent currents. We consider a system of length $L=96\mu m$ and work at fixed total number of $N = 3\times 10^5$ $^{23}$Na atoms.

In order to have a finite current in component 1, we impose a linearly varying phase of the order parameter $\psi_1=\sqrt{n_1}\exp(i\phi_1)$, which satisfies the proper boundary condition, i.e., $\phi_1(x) =W2\pi x/L$ for $0\le x \le L$, with $W=\pm1,\pm 2, ..$ being the winding number.
The imprinting of such a linear phase will generate a state with the desired permanent current $j_1=n_1 v_1$ with $v_1=\hbar\partial_x\phi_1/m= W 2\pi\hbar/mL $.

We study the sound excitations by adding a static, low-amplitude potential in the form of 
\begin{equation}\label{eq:prot_pot}
    V_{ext}(x) = \lambda\cos (qx)
\end{equation} 
acting on both components. The perturbation wavevector $q=2\pi / L$ excites sound waves, provided that $q$ is sufficiently small, and thus belongs to the linear part of the dispersion law. Our protocol begins with the preparation of the initial state in the presence of this potential, by means of imaginary time evolution of Eq. (\ref{eq:2c_gpe}). Then, at the beginning of the real-time evolution ($t\ge 0$), we suddenly remove it to trigger the density profile dynamics. By measuring the two single-component densities, we gain access to the lowest-energy modes propagating in the system.  A similar protocol was used to experimentally measure the Doppler effect in a single-component BEC \cite{Kumar2016}. We have also recently applied it  to dipolar condensates to study the speed of sound \cite{Sindik2023SoundSA} and the Doppler effect \cite{dopplerSupersolid} across the superfluid-to-supersolid phase transition.

According to linear response theory, the time dependent averages
\begin{equation}\label{eq:signal}
    \langle \cos(qx)\rangle_i =  \int dx\ n_i(x,t) \cos (qx) \; ,
\end{equation}
for $t\ge 0$, give access to the frequencies of the phonon modes excited by the periodic density perturbation. For a mixture at rest, this protocol yields the excitation of two modes, one associated with the density-like oscillation of the mixture and one with the spin-like one. In the presence of a finite current, each mode splits into two, due to the Doppler effect. Since, for small wave vectors $q$ the phonon dispersion relation is linear, the frequency of the excited modes gives access to the corresponding speed of sound $c_i= \omega_i/q$. Finally, for each sound, the Doppler shifts are calculated as $\Delta c_i = (\omega_i^+ - \omega_i^- )/2q$, where $\omega_i^+ $ and $\omega_i^- $ represent, respectively, the frequency of the co- and counter-propagating phonons. The numerical results of this protocol are shown in Fig.~(\ref{fig:fig1}a) and Fig.~(\ref{fig:fig1}b). As anticipated, the agreement with the hydrodynamic model is very good.

However, the numerical procedure presented in this section is not well suited for an experimental observation of the Doppler effect, as it requires very long time evolution to accurately discriminate the Doppler split frequencies of each mode. In the next section, we  propose an improved way to measure the Doppler effect on a short time-scale.

% End of the technical part

\subsection{Selective mode measurement}\label{sec:protocol}

In order to better identify the density-like and the spin-like modes, we modify the protocol by applying the potential (\ref{eq:prot_pot}) with different amplitudes $\lambda_i$ to each component. We then calculate the density response by solving the HD equations
\begin{subequations}\label{eq:ext_prot_hd_eq}
\begin{align}
    \pt n_1+\px (n_1v_1)&=0 \label{eq:ep_1} \\
    \pt n_2+\px (n_2v_2)&=0 \label{eq:ep_2} \\
    \pt v_1 + \px \left(\frac{1}{2}v_1^2 + \mu_1 + \lambda_1e^{-i(\omega t - qx)}\right) &=0 \label{eq:ep_3}\\
    \pt v_2 + \px \left(\frac{1}{2}v_2^2 + \mu_2 + \lambda_2e^{-i(\omega t - qx)}\right) &=0 \label{eq:ep_4},
\end{align}
\end{subequations}
where $n_i,\ v_i,\ \mu_i$ are, respectively, the density, velocity, and chemical potential of the $i$-th component. The time-dependent perturbation $\lambda_i\exp(i(qx-\omega t))$ is assumed to slowly vary in space and time. For simplicity, we consider only systems in the absence of currents. The results of this section are expected to be marginally affected by the presence of  small stationary currents.

We use the linearization method to calculate the total density response $\delta n$ to the two perturbations $\lambda_i$
\begin{equation}\label{eq:deltan}
    \delta n(\omega,q)\!=\! 
    \frac
    {\displaystyle 
\sum_{k=1,2}\lambda_kn^0_k\left[ \omega^{2} + q^2n^0_{3-k}(g_{12}  - g_{(3-k)(3-k)}) \right]q^2  }
    {\left[ \omega^{2} -(qc_d^0)^2 \right] \left[ \omega^{2} -(qc_s^0)^2 \right]},
\end{equation}
where the two speeds of sound $c_s^0$ and $c_d^0$ are given by (\ref{eq:c0s}). As anticipated, the density response has in general two pairs of poles - one pair for every sound mode. To decouple one of them, the numerator of (\ref{eq:deltan}) must take the form $(n_1^0\lambda_1 + n_2^0\lambda_2)\left[ \omega^2 - (qc_i^{0})^2 \right]q^2$ to cancel out the two corresponding poles. This condition introduces a relation between the two amplitudes $\lambda_i$, which reads:
\begin{equation}\label{eq:l1l2}
    \frac{\lambda_1}{\lambda_2} = - \frac{g_{11}-g_{12} - \left(c_i^0\right)^2/n^0_1}{g_{22}-g_{12} - \left(c_i^0\right)^2/n^0_2},
\end{equation}
where the index $i=d,s$ indicates the mode selected to be suppressed. The above condition defines a ratio of the two external perturbation strengths applied to the two components of the BEC mixture, provided that their densities $n_i$ or their intraspecies coupling constant $g_{ii}$ are different. In the symmetric case, i.e. $g_{11}=g_{22}\neq g_{12}$ and $n_1^0 = n_2^0$, it can easily be shown that equations (\ref{eq:ext_prot_hd_eq}) require $\lambda_1 = \lambda_2$ ($\lambda_1 = -\lambda_2$) for the spin-like (density-like) mode to vanish from the observed signal.

\subsection{Precession}\label{sec:precession}
The protocol proposed in this work enables the measurement of frequencies of all modes excited with the initial perturbation. In the context of the Doppler effect, this information is redundant, as we only need to know the difference between the two Doppler-split modes. Direct access to this quantity is obtained by studying the precession of the density profile during the time evolution. This approach has already been applied in a single-component BEC \cite{Kumar2016}. Using the protocol with the appropriate choice (\ref{eq:l1l2}) of the amplitudes $\lambda_i$, we show that the Doppler shift of each sound can be efficiently studied by measuring the precession of the density profile.

According to linear response theory, the time evolution of the density variation $\delta n (x,t)$ induced by a small perturbation is given by all the modes propagating in the system. Provided that the perturbation strengths satisfy the condition (\ref{eq:l1l2}), in the presence of a small current, there are only two Doppler-split modes affecting the total density:
\begin{equation}\label{eq:dnxt}
    \delta n (x,t) = A_i^+ \cos \left( q x -  \omega_i^+ t\right) + A_i^- \cos \left( q x +  \omega_i^- t\right),
\end{equation}
where $\omega^\pm = \omega_i^0 \pm \Omega_i$, and $\Omega_i$ is the Doppler shift of the $i$-th mode. For small currents, the difference between the two amplitudes $A_i^\pm$ can be neglected, as $A_i^+ - A_i^- \propto  \Omega_i/[(\omega_i^0)^2 - \Omega_i^2]$. Under this assumption, the density profile evolves as 
\begin{equation}\label{eq:dnxt2}
    \delta n (x,t) \approx A_i\cos (qx - \Omega_it)\cos (\omega_i^0 t),
\end{equation}
where $A_i=A_i^++A_i^-$ and the Doppler shift is explicitly separated from the rest value of the frequency of the  $i$-th mode $\omega_i^0$. At stroboscopic times, when $\omega_i^0 t$ equals an integer multiplication of $2\pi$, the position of the density maximum changes with respect to its initial value by $\Omega_i t$. Thus, the slope of this dependence provides the value of the Doppler shift. 

\begin{figure}[!ht]
    \centering
    \includegraphics[width=0.995\linewidth]{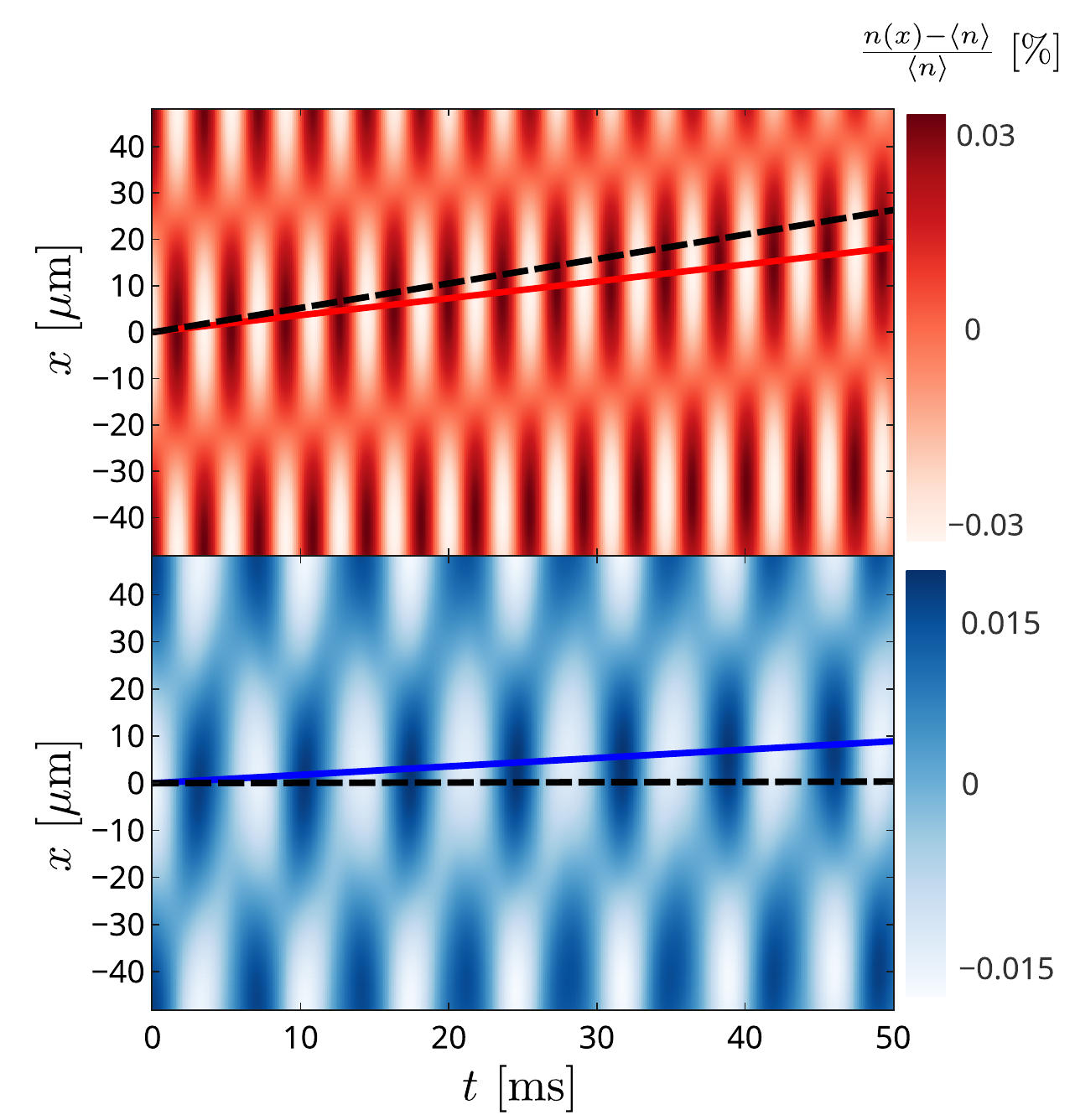}
    \caption{Precession of the density profile $n(x)-\langle n\rangle$ with selectively excited density-like mode (upper panel) and spin-like mode (lower panel). In each panel, the solid line indicates the trajectory of the density maximum at stroboscopic times, as given by the hydrodynamic formula (\ref{eq:gpe_doppler}). The results correspond to the following gas parameters: $m_z=0.1$, $g_{ds}/g = 0.1$, $g_{12}/g=0.55$ and $v_T^0 = w^0 = 3\times 2\pi \hbar /(mL)$, for which $\delta_d = 0.67$ and $\delta_s = 0.33$.  The dashed line presents the Doppler shift for the same gas parameters but with interspecies interaction $g_{12}=0$, where the density-like (spin-like) mode corresponds to the oscillation of the density of the first (second) component.   }
    \label{fig:precession}
\end{figure}

To demonstrate the potential of this method, we simulate the two-component Bose mixture in the same one-dimensional configuration as in the previous section.  To amplify the signal, we imprint a current with the winding number $W=3$ in the phase pattern of the first component, while the second remains at rest. To independently study the Doppler shifts of the two sounds, we fix the protocol amplitudes according to equation (\ref{eq:l1l2}). At time $t=0$ we suddenly remove the spin dependent potential to observe the time evolution of the total density profile $n(x,t)$. In Fig.~(\ref{fig:precession}) we present the results of the numerical simulation for the precession for both sound modes. The maximum of the total density clearly follows the hydrodynamic prediction indicated by the corresponding solid line and thus reveals the importance of the interspecies interaction.

\section{Andreev-Bashkin Effect}
In the previous sections, we have considered the mean-field case, where the superfluid current in each component is determined only by its superfluid velocity. In general, however, cross terms are present due to the so-called Andreev-Bashkin dissipationless drag \cite{ABeffect}.  As a result, a finite velocity field of one component gives a contribution to the current in the other superfluid component:
\begin{eqnarray}
   j_i = (n_i-n_D)v_i+n_D v_{3-i}
\end{eqnarray} 
where $n_D$ is the so-called superfluid drag and can be seen as the (quantum) dressing of one component by the other.  The current entrainment introduces an additional term in the total energy density $\epsilon_D = - \frac{1}{2}n_D(v_1-v_2)^2 = -2n_{D} w^2$, which alters the hydrodynamic equations:
\begin{subequations}\label{eq:hdAB}
\begin{align}
    \pt n+\partial_x j&=0 \\
    \pt s_z+\partial_x( j_z-4n_D w)&=0  \\
    \pt v_T+\partial_x\left( \frac{v_T^2}{2}+\frac{ w^2}{2} -2\partial_n n_D w^2+\frac{\mu}{m}\right)&=0\\
    \pt w+\partial_x\left( wv_T-
    2\partial_{s_z} n_D w^2
    +\frac{h}{m}\right)&=0.
\end{align}     
\end{subequations}
Linearization of these equations leads to
the two speeds of sound at rest:
\begin{widetext}
\begin{equation}
%c_{d/s}^0 = \left[\frac{n^0g + s_z^0g_{ds} -4n^0_Dg_s}{2m} 
%             \pm \frac{1}{2m}\sqrt{(n^0g + s_z^0g_{ds} -4n_D^0g_s)^2 - \Delta \left[(n^0)^2-4n_D^0n^0-(s_z^0)^2\right]}   \right]^{1/2}
c_{d/s}^0 = \left[\frac{n^0}{2m} \left(g + m_zg_{ds} -4f_Dg_s
             \pm \sqrt{(g + m_zg_{ds} -4f_Dg_s)^2 - \Delta \left[1-4f_D-m_z^2\right]}   \right)\right]^{1/2},
\end{equation}
\end{widetext}
where we introduce the drag fraction $f_D = n_D^0/n^0$. Linearization of the hydrodynamic equations (\ref{eq:hdAB}) yields the same form of the Doppler shift of the two sound modes: $\Delta c_{d/s}= v_T^0 + \delta_{d/s}w^0$, where the $\delta_{d/s}$ term is modified by the Andreev-Bashkin drag and reads:
\begin{widetext}
\begin{equation}\label{eq:delta_AB}
    %\delta_{d} =-\delta_s=  \frac{gs_z^0 +g_{ds}(n^0-2n_D^0)-4\partial_{s_z} n_D^0 \left[ \left(c^0_{d/s}\right)^2 - \mu^0\right] - 4\partial_n n_D^0h^0 }{\sqrt{\left(n^0g + s_z^0g_{ds} -4n_D^0g_s\right)^2 - \Delta \left[(n^0)^2-4n_D^0n^0-(s_z^0)^2\right]} }  ,
    \delta_{d} =-\delta_s=  \frac{gm_z +g_{ds}(1-2f_D)-\frac{4}{n^0}\partial_{s_z} n_D^0 \left[ \left(c^0_{d/s}\right)^2 - \mu^0\right] - \frac{4}{n^0}\partial_n n_D^0h^0 }{\sqrt{\left(g + m_zg_{ds} -4f_Dg_s\right)^2 - \Delta \left[1-4f_D-m_z^2\right]} }  ,
\end{equation}
\end{widetext}
where $\mu^0 = g_dn^0 + g_{ds}s_z^0/2$ and $h^0=g_ss_z^0 + g_{ds}n^0/2$. For weakly interacting Bose mixtures, the drag density $n_D$ is very small, because of its linear dependence on the two gas parameters $\eta_i = \sqrt{n_ia_{ii}^3}$ \cite{shevchenko2007}. Thus, inclusion of this effect will not visibly change the Doppler shift. Nonetheless, it is interesting to notice a qualitative change in the large imbalance regime $s_z \approx -n$. There, the drag density is approximately equal:
\begin{equation}
    %n_D \approx 0.8 n_1 \frac{a_{12}^2}{a_{22}^2}\sqrt{n_2a_{22}^3},
    n_D \approx 0.4 (n + s_z) \left(\frac{g_{12}}{g-g_{ds}}\right)^2\eta_2
\end{equation}
and the two derivatives read:
\begin{subequations}
\begin{align}
    %\partial_{s_z}n_D & = 0.2 \left(\frac{a_{12}}{a_{22}}\right)^2\sqrt{n_2a_{22}^3} \lt 2 - \frac{n_1}{n_2} \rt \\
    %\partial_{n}n_D & = 0.2 \left(\frac{a_{12}}{a_{22}}\right)^2\sqrt{n_2a_{22}^3} \lt 2 + \frac{n_1}{n_2} \rt 
    \partial_{s_z}n_D & = 0.2 \left(\frac{g_{12}}{g - g_{ds}}\right)^2 \lt 2 - \frac{1+m_z}{1-m_z} \rt \eta_2 \\
    \partial_{n}n_D & = 0.2 \left(\frac{g_{12}}{g - g_{ds}}\right)^2 \lt 2 + \frac{1+m_z}{1-m_z} \rt \eta_2
\end{align}
\end{subequations}
This result shows that, in contrast to the mean-field case, for $m_z=-g_{ds}/g$, the numerator of (\ref{eq:delta_AB}) does not vanish. Consequently, for this value of magnetization, the presence of the dissipationless drag lifts the independence of the Doppler shift from the $g_{12}$ coupling constant. While the effect is negligible in dilute Bose mixtures, in other systems, the Doppler effect may serve as a good observable for detecting the Andreev-Bashkin effect. 

% Refs?:
% - AB enhanced in Bose-Hubbard model \cite{AlessioLinearResponseAB}
% - AB enhanced in 1D BEC using MC. \cite{Stefano1DBEC_AB} How is that different from our mixtures?
% 

%In the parameter regime close to the $\mathbb{Z}_2$ symmetric case, the drag density is approximately equal:
% \begin{equation}
%     n_D \approx0.2\frac{n^2-s_z^2}{n}\frac{a^2_{12}}{a^2}\sqrt{na^3}.
% \end{equation}
% In such a case, the two derivatives $\partial_{s_z}n_D^0$ and $\partial_{n}n_D^0$ vanish, and the Doppler effect becomes unaffected by the Andreev-Bashkin effect. However, in the opposite limit, when $s_z \approx -n$, the result is qualitatively different. In this regime, the drag density is approximately equal:

\section{Conclusions}
In this work, we investigated the Doppler effect in binary Bose mixtures, where we revealed a non-trivial sound entrainment caused by the intraspecies interaction and the relative velocity between the two components. Using the hydrodynamic theory of two coupled superfluids, we derived a formula describing the Doppler shift of the spin-like and the density-like sounds. The complex dependence of the Doppler shift on the thermodynamic properties of the system has already been the subject of previous theoretical studies \cite{dopplerSupersolid,NepoHe34,NepoHeII,Khalatnikov56,NepoHe4thSound}. Among all these platforms, the Bose mixture gas seems to be the most promising one in the context of an experimental observation of the non-trivial Doppler effect, due to the versatile access to both wave functions and their long lifetime. Furthermore, employing the linear response theory, we proposed an experimental protocol to individually study the Doppler shift of each sound mode. Our proposal generalizes the method that was successfully applied in a single-component BEC \cite{Kumar2016} to observe the same effect.  

Finally, we considered the Andreev-Bashkin effect in the context of the Doppler effect. We demonstrated that, in weakly coupled Bose mixtures, this effect slightly alters the Doppler shift of each sound. In principle, in systems where the Andreev-Bashkin effect is more pronounced, our protocol provides the means to measure the dissipation-less drag.

Lastly, we stress that the results of this work are general to all superfluid mixtures, being applicable to fermionic gases, as well as mixtures of different quantum statistics.

We acknowledge useful discussions with Gabriele Ferrari, Giacomo Lamporesi and Marija \v{S}indik. This work has been supported by the Provincia Autonoma di Trento, CINECA consortium through the award under the ISCRA initiative  for the availability of HPC resources. Part of the work was computed on ``Deeplearning cluster" supported by the initiative ``Dipartimenti di Eccellenza 2018-2022 (Legge 232/2016)" funded by the MUR.

\textit{Data availability statement} - The data that support the findings of this article are not publicly available upon publication because it is not technically feasible and/or the cost of preparing, depositing, and hosting the data would be prohibitive within the terms of this research project. The data are available from the authors upon reasonable request.

\bibliography{bibliomixture}

\begin{thebibliography}{22}%
\makeatletter
\providecommand \@ifxundefined [1]{%
 \@ifx{#1\undefined}
}%
\providecommand \@ifnum [1]{%
 \ifnum #1\expandafter \@firstoftwo
 \else \expandafter \@secondoftwo
 \fi
}%
\providecommand \@ifx [1]{%
 \ifx #1\expandafter \@firstoftwo
 \else \expandafter \@secondoftwo
 \fi
}%
\providecommand \natexlab [1]{#1}%
\providecommand \enquote  [1]{``#1''}%
\providecommand \bibnamefont  [1]{#1}%
\providecommand \bibfnamefont [1]{#1}%
\providecommand \citenamefont [1]{#1}%
\providecommand \href@noop [0]{\@secondoftwo}%
\providecommand \href [0]{\begingroup \@sanitize@url \@href}%
\providecommand \@href[1]{\@@startlink{#1}\@@href}%
\providecommand \@@href[1]{\endgroup#1\@@endlink}%
\providecommand \@sanitize@url [0]{\catcode `\\12\catcode `\$12\catcode
  `\&12\catcode `\#12\catcode `\^12\catcode `\_12\catcode `\%12\relax}%
\providecommand \@@startlink[1]{}%
\providecommand \@@endlink[0]{}%
\providecommand \url  [0]{\begingroup\@sanitize@url \@url }%
\providecommand \@url [1]{\endgroup\@href {#1}{\urlprefix }}%
\providecommand \urlprefix  [0]{URL }%
\providecommand \Eprint [0]{\href }%
\providecommand \doibase [0]{https://doi.org/}%
\providecommand \selectlanguage [0]{\@gobble}%
\providecommand \bibinfo  [0]{\@secondoftwo}%
\providecommand \bibfield  [0]{\@secondoftwo}%
\providecommand \translation [1]{[#1]}%
\providecommand \BibitemOpen [0]{}%
\providecommand \bibitemStop [0]{}%
\providecommand \bibitemNoStop [0]{.\EOS\space}%
\providecommand \EOS [0]{\spacefactor3000\relax}%
\providecommand \BibitemShut  [1]{\csname bibitem#1\endcsname}%
\let\auto@bib@innerbib\@empty
%</preamble>
\bibitem [{\citenamefont {Khalatnikov}()}]{Khalatnikov56}%
  \BibitemOpen
  \bibfield  {author} {\bibinfo {author} {\bibfnamefont {I.}~\bibnamefont
  {Khalatnikov}},\ }\bibfield  {title} {\bibinfo {title} {The propagation of
  sound in moving helium ii and the effect of a thermal current upon the
  propagation of second sound},\ }\href@noop {} {\bibinfo  {journal} {Zh. Eksp.
  Teor. Fiz 30, 617 (1956) [Sov. Phys. JETP 3, 649 (1956)]}\ }\BibitemShut
  {NoStop}%
\bibitem [{\citenamefont {Nepomnyashchy}(1993)}]{NepoHeII}%
  \BibitemOpen
\bibfield  {journal} {  }\bibfield  {author} {\bibinfo {author} {\bibfnamefont
  {Y.~A.}\ \bibnamefont {Nepomnyashchy}},\ }\bibfield  {title} {\bibinfo
  {title} {Unusual doppler effect in he ii},\ }\href
  {https://doi.org/10.1103/PhysRevB.47.905} {\bibfield  {journal} {\bibinfo
  {journal} {Phys. Rev. B}\ }\textbf {\bibinfo {volume} {47}},\ \bibinfo
  {pages} {905} (\bibinfo {year} {1993})}\BibitemShut {NoStop}%
\bibitem [{\citenamefont {Nepomnyashchy}\ and\ \citenamefont
  {Revzen}(1991)}]{NepoHe4thSound}%
  \BibitemOpen
  \bibfield  {author} {\bibinfo {author} {\bibfnamefont {Y.}~\bibnamefont
  {Nepomnyashchy}}\ and\ \bibinfo {author} {\bibfnamefont {M.}~\bibnamefont
  {Revzen}},\ }\bibfield  {title} {\bibinfo {title} {Curious doppler shift of
  fourth sound in the low temperature limit},\ }\href
  {https://doi.org/https://doi.org/10.1016/0375-9601(92)90770-M} {\bibfield
  {journal} {\bibinfo  {journal} {Physics Letters A}\ }\textbf {\bibinfo
  {volume} {161}},\ \bibinfo {pages} {164} (\bibinfo {year}
  {1991})}\BibitemShut {NoStop}%
\bibitem [{\citenamefont {Gov}\ \emph {et~al.}(1993)\citenamefont {Gov},
  \citenamefont {Mann}, \citenamefont {Nepomnyashchy},\ and\ \citenamefont
  {Revzen}}]{NepoHe34}%
  \BibitemOpen
  \bibfield  {author} {\bibinfo {author} {\bibfnamefont {N.}~\bibnamefont
  {Gov}}, \bibinfo {author} {\bibfnamefont {A.}~\bibnamefont {Mann}}, \bibinfo
  {author} {\bibfnamefont {Y.}~\bibnamefont {Nepomnyashchy}},\ and\ \bibinfo
  {author} {\bibfnamefont {M.}~\bibnamefont {Revzen}},\ }\bibfield  {title}
  {\bibinfo {title} {Unusual doppler shift of fourth sound in a $^{3}${H}e -
  $^{4}${H}e mixture},\ }\href
  {https://doi.org/https://doi.org/10.1016/0375-9601(93)90069-C} {\bibfield
  {journal} {\bibinfo  {journal} {Physics Letters A}\ }\textbf {\bibinfo
  {volume} {182}},\ \bibinfo {pages} {149} (\bibinfo {year}
  {1993})}\BibitemShut {NoStop}%
\bibitem [{\citenamefont {Zawi\ifmmode~\acute{s}\else \'{s}\fi{}lak}\ \emph
  {et~al.}(2025)\citenamefont {Zawi\ifmmode~\acute{s}\else \'{s}\fi{}lak},
  \citenamefont {\ifmmode~\check{S}\else \v{S}\fi{}indik}, \citenamefont
  {Stringari},\ and\ \citenamefont {Recati}}]{dopplerSupersolid}%
  \BibitemOpen
  \bibfield  {author} {\bibinfo {author} {\bibfnamefont {T.}~\bibnamefont
  {Zawi\ifmmode~\acute{s}\else \'{s}\fi{}lak}}, \bibinfo {author}
  {\bibfnamefont {M.}~\bibnamefont {\ifmmode~\check{S}\else \v{S}\fi{}indik}},
  \bibinfo {author} {\bibfnamefont {S.}~\bibnamefont {Stringari}},\ and\
  \bibinfo {author} {\bibfnamefont {A.}~\bibnamefont {Recati}},\ }\bibfield
  {title} {\bibinfo {title} {Anomalous {D}oppler effect in superfluid and
  supersolid atomic gases},\ }\href
  {https://doi.org/10.1103/PhysRevLett.134.226001} {\bibfield  {journal}
  {\bibinfo  {journal} {Phys. Rev. Lett.}\ }\textbf {\bibinfo {volume} {134}},\
  \bibinfo {pages} {226001} (\bibinfo {year} {2025})}\BibitemShut {NoStop}%
\bibitem [{\citenamefont {Saslow}(2025)}]{saslowInertia}%
  \BibitemOpen
  \bibfield  {author} {\bibinfo {author} {\bibfnamefont {W.~M.}\ \bibnamefont
  {Saslow}},\ }\href {https://arxiv.org/abs/2501.06338} {\bibinfo {title}
  {Dynamics of supersolid state: normal fluid, superfluid, and supersolid
  velocities}} (\bibinfo {year} {2025}),\ \Eprint
  {https://arxiv.org/abs/2501.06338} {arXiv:2501.06338 [cond-mat.quant-gas]}
  \BibitemShut {NoStop}%
\bibitem [{\citenamefont {Andreev}\ and\ \citenamefont
  {Bashkin}(1975)}]{ABeffect}%
  \BibitemOpen
  \bibfield  {author} {\bibinfo {author} {\bibfnamefont {A.~F.}\ \bibnamefont
  {Andreev}}\ and\ \bibinfo {author} {\bibfnamefont {E.~P.}\ \bibnamefont
  {Bashkin}},\ }\bibfield  {title} {\bibinfo {title} {Three-velocity
  hydrodynamics of superfluid solutions},\ }\href
  {https://www.osti.gov/biblio/4106539} {\bibfield  {journal} {\bibinfo
  {journal} {Zh. Eksp. Teor. Fiz., v. 69, no. 1, pp. 319-326}\ } (\bibinfo
  {year} {1975})}\BibitemShut {NoStop}%
\bibitem [{Note1()}]{Note1}%
  \BibitemOpen
  \bibinfo {note} {The stability in terms of the intra- and interspecies
  interactions reads: $g_1 g_2-g_{12}^2\ge 0$. For $g_{12}<0$ the instability
  would lead to the collapse of the mixture.}\BibitemShut {Stop}%
\bibitem [{\citenamefont {Pitaevskii}\ and\ \citenamefont
  {Stringari}(2016)}]{BecBook2016}%
  \BibitemOpen
  \bibfield  {author} {\bibinfo {author} {\bibfnamefont {L.}~\bibnamefont
  {Pitaevskii}}\ and\ \bibinfo {author} {\bibfnamefont {S.}~\bibnamefont
  {Stringari}},\ }\href@noop {} {\emph {\bibinfo {title} {{B}ose-{E}instein
  condensation and superfluidity}}}\ (\bibinfo  {publisher} {Oxford University
  Press},\ \bibinfo {year} {2016})\BibitemShut {NoStop}%
\bibitem [{\citenamefont {Armaitis}\ \emph {et~al.}(2015)\citenamefont
  {Armaitis}, \citenamefont {Stoof},\ and\ \citenamefont
  {Duine}}]{DuineHD2BEC}%
  \BibitemOpen
  \bibfield  {author} {\bibinfo {author} {\bibfnamefont {J.}~\bibnamefont
  {Armaitis}}, \bibinfo {author} {\bibfnamefont {H.~T.~C.}\ \bibnamefont
  {Stoof}},\ and\ \bibinfo {author} {\bibfnamefont {R.~A.}\ \bibnamefont
  {Duine}},\ }\bibfield  {title} {\bibinfo {title} {Hydrodynamic modes of
  partially condensed bose mixtures},\ }\href
  {https://doi.org/10.1103/PhysRevA.91.043641} {\bibfield  {journal} {\bibinfo
  {journal} {Phys. Rev. A}\ }\textbf {\bibinfo {volume} {91}},\ \bibinfo
  {pages} {043641} (\bibinfo {year} {2015})}\BibitemShut {NoStop}%
\bibitem [{Note2()}]{Note2}%
  \BibitemOpen
  \bibinfo {note} {In the single-component notation, the two speeds of sound
  have the familiar form: $c_{d/s}^0 = \protect \frac {1}{\protect \sqrt {2m}}
  \left [g_{11}n_1 + g_{22}n_2 \pm \protect \sqrt {(g_{11}n_1 - g_{22}n_2)^2 +
  4n_1n_2g_{12}^2}\right ]^{\protect \frac {1}{2}} $}\BibitemShut {NoStop}%
\bibitem [{\citenamefont {Cominotti}\ \emph {et~al.}(2022)\citenamefont
  {Cominotti}, \citenamefont {Berti}, \citenamefont {Farolfi}, \citenamefont
  {Zenesini}, \citenamefont {Lamporesi}, \citenamefont {Carusotto},
  \citenamefont {Recati},\ and\ \citenamefont {Ferrari}}]{Gabriele2BECsound}%
  \BibitemOpen
  \bibfield  {author} {\bibinfo {author} {\bibfnamefont {R.}~\bibnamefont
  {Cominotti}}, \bibinfo {author} {\bibfnamefont {A.}~\bibnamefont {Berti}},
  \bibinfo {author} {\bibfnamefont {A.}~\bibnamefont {Farolfi}}, \bibinfo
  {author} {\bibfnamefont {A.}~\bibnamefont {Zenesini}}, \bibinfo {author}
  {\bibfnamefont {G.}~\bibnamefont {Lamporesi}}, \bibinfo {author}
  {\bibfnamefont {I.}~\bibnamefont {Carusotto}}, \bibinfo {author}
  {\bibfnamefont {A.}~\bibnamefont {Recati}},\ and\ \bibinfo {author}
  {\bibfnamefont {G.}~\bibnamefont {Ferrari}},\ }\bibfield  {title} {\bibinfo
  {title} {Observation of massless and massive collective excitations with
  faraday patterns in a two-component superfluid},\ }\href
  {https://doi.org/10.1103/PhysRevLett.128.210401} {\bibfield  {journal}
  {\bibinfo  {journal} {Phys. Rev. Lett.}\ }\textbf {\bibinfo {volume} {128}},\
  \bibinfo {pages} {210401} (\bibinfo {year} {2022})}\BibitemShut {NoStop}%
\bibitem [{\citenamefont {Kim}\ \emph {et~al.}(2020)\citenamefont {Kim},
  \citenamefont {Hong},\ and\ \citenamefont {Shin}}]{Shin2BECsound}%
  \BibitemOpen
  \bibfield  {author} {\bibinfo {author} {\bibfnamefont {J.~H.}\ \bibnamefont
  {Kim}}, \bibinfo {author} {\bibfnamefont {D.}~\bibnamefont {Hong}},\ and\
  \bibinfo {author} {\bibfnamefont {Y.}~\bibnamefont {Shin}},\ }\bibfield
  {title} {\bibinfo {title} {Observation of two sound modes in a binary
  superfluid gas},\ }\href {https://doi.org/10.1103/PhysRevA.101.061601}
  {\bibfield  {journal} {\bibinfo  {journal} {Phys. Rev. A}\ }\textbf {\bibinfo
  {volume} {101}},\ \bibinfo {pages} {061601} (\bibinfo {year}
  {2020})}\BibitemShut {NoStop}%
\bibitem [{\citenamefont {Piekarski}\ \emph {et~al.}(2025)\citenamefont
  {Piekarski}, \citenamefont {Cherroret}, \citenamefont {Aladjidi},\ and\
  \citenamefont {Glorieux}}]{piekarski2BEC}%
  \BibitemOpen
  \bibfield  {author} {\bibinfo {author} {\bibfnamefont {C.}~\bibnamefont
  {Piekarski}}, \bibinfo {author} {\bibfnamefont {N.}~\bibnamefont
  {Cherroret}}, \bibinfo {author} {\bibfnamefont {T.}~\bibnamefont
  {Aladjidi}},\ and\ \bibinfo {author} {\bibfnamefont {Q.}~\bibnamefont
  {Glorieux}},\ }\bibfield  {title} {\bibinfo {title} {Spin and density modes
  in a binary fluid of light},\ }\href {https://doi.org/10.1103/s58b-3mmx}
  {\bibfield  {journal} {\bibinfo  {journal} {Phys. Rev. Lett.}\ }\textbf
  {\bibinfo {volume} {134}},\ \bibinfo {pages} {223403} (\bibinfo {year}
  {2025})}\BibitemShut {NoStop}%
\bibitem [{\citenamefont {Bienaim\'e}\ \emph {et~al.}(2016)\citenamefont
  {Bienaim\'e}, \citenamefont {Fava}, \citenamefont {Colzi}, \citenamefont
  {Mordini}, \citenamefont {Serafini}, \citenamefont {Qu}, \citenamefont
  {Stringari}, \citenamefont {Lamporesi},\ and\ \citenamefont
  {Ferrari}}]{spindipole2016}%
  \BibitemOpen
  \bibfield  {author} {\bibinfo {author} {\bibfnamefont {T.}~\bibnamefont
  {Bienaim\'e}}, \bibinfo {author} {\bibfnamefont {E.}~\bibnamefont {Fava}},
  \bibinfo {author} {\bibfnamefont {G.}~\bibnamefont {Colzi}}, \bibinfo
  {author} {\bibfnamefont {C.}~\bibnamefont {Mordini}}, \bibinfo {author}
  {\bibfnamefont {S.}~\bibnamefont {Serafini}}, \bibinfo {author}
  {\bibfnamefont {C.}~\bibnamefont {Qu}}, \bibinfo {author} {\bibfnamefont
  {S.}~\bibnamefont {Stringari}}, \bibinfo {author} {\bibfnamefont
  {G.}~\bibnamefont {Lamporesi}},\ and\ \bibinfo {author} {\bibfnamefont
  {G.}~\bibnamefont {Ferrari}},\ }\bibfield  {title} {\bibinfo {title}
  {Spin-dipole oscillation and polarizability of a binary bose-einstein
  condensate near the miscible-immiscible phase transition},\ }\href
  {https://doi.org/10.1103/PhysRevA.94.063652} {\bibfield  {journal} {\bibinfo
  {journal} {Phys. Rev. A}\ }\textbf {\bibinfo {volume} {94}},\ \bibinfo
  {pages} {063652} (\bibinfo {year} {2016})}\BibitemShut {NoStop}%
\bibitem [{Note3()}]{Note3}%
  \BibitemOpen
  \bibinfo {note} {If $g_{12}$ were negative $\pm $ in the Eq. (\ref {eq:c0s})
  has to be replaced by $\mp $ .}\BibitemShut {Stop}%
\bibitem [{Note4()}]{Note4}%
  \BibitemOpen
  \bibinfo {note} {Notice that $j_z$ is the $z$-component of the Noether SU(2)
  current \begin {equation} \protect \mathbf {j}_s=v\protect \mathbf
  {s}+\protect \frac {\hbar }{2m}\left (\protect \frac {\protect \mathbf
  {s}}{n}\times \partial _x \protect \mathbf {s}\right ), \end {equation} where
  the first term corresponds to the spin advection, while the second is the
  so-called quantum torque. In the case of a SU(2) symmetric atomic mixtures,
  i.e., $g_s=g_{ds}=0$, the full spin would be conserved, and spin HD reduces
  to Noether conservation law: $\partial _t\protect \mathbf {s}+\partial
  _x\protect \mathbf {j}_s=0 $.}\BibitemShut {Stop}%
\bibitem [{Note5()}]{Note5}%
  \BibitemOpen
  \bibinfo {note} {We determined the four solutions as the roots of the
  fourth-order polynomial, resulting from the determinant of Eq. (\ref
  {eq:matrix}).}\BibitemShut {Stop}%
\bibitem [{\citenamefont {Stringari}(1996)}]{stringari96}%
  \BibitemOpen
  \bibfield  {author} {\bibinfo {author} {\bibfnamefont {S.}~\bibnamefont
  {Stringari}},\ }\bibfield  {title} {\bibinfo {title} {Collective excitations
  of a trapped {B}ose-condensed gas},\ }\href
  {https://doi.org/10.1103/PhysRevLett.77.2360} {\bibfield  {journal} {\bibinfo
   {journal} {Phys. Rev. Lett.}\ }\textbf {\bibinfo {volume} {77}},\ \bibinfo
  {pages} {2360} (\bibinfo {year} {1996})}\BibitemShut {NoStop}%
\bibitem [{\citenamefont {Kumar}\ \emph {et~al.}(2016)\citenamefont {Kumar},
  \citenamefont {Anderson}, \citenamefont {Phillips}, \citenamefont {Eckel},
  \citenamefont {Campbell},\ and\ \citenamefont {Stringari}}]{Kumar2016}%
  \BibitemOpen
  \bibfield  {author} {\bibinfo {author} {\bibfnamefont {A.}~\bibnamefont
  {Kumar}}, \bibinfo {author} {\bibfnamefont {N.}~\bibnamefont {Anderson}},
  \bibinfo {author} {\bibfnamefont {W.~D.}\ \bibnamefont {Phillips}}, \bibinfo
  {author} {\bibfnamefont {S.}~\bibnamefont {Eckel}}, \bibinfo {author}
  {\bibfnamefont {G.~K.}\ \bibnamefont {Campbell}},\ and\ \bibinfo {author}
  {\bibfnamefont {S.}~\bibnamefont {Stringari}},\ }\bibfield  {title} {\bibinfo
  {title} {Minimally destructive, doppler measurement of a quantized flow in a
  ring-shaped {B}ose--{E}instein condensate},\ }\href
  {https://doi.org/10.1088/1367-2630/18/2/025001} {\bibfield  {journal}
  {\bibinfo  {journal} {New Journal of Physics}\ }\textbf {\bibinfo {volume}
  {18}},\ \bibinfo {pages} {025001} (\bibinfo {year} {2016})}\BibitemShut
  {NoStop}%
\bibitem [{\citenamefont {\ifmmode~\check{S}\else \v{S}\fi{}indik}\ \emph
  {et~al.}(2024)\citenamefont {\ifmmode~\check{S}\else \v{S}\fi{}indik},
  \citenamefont {Zawi\ifmmode~\acute{s}\else \'{s}\fi{}lak}, \citenamefont
  {Recati},\ and\ \citenamefont {Stringari}}]{Sindik2023SoundSA}%
  \BibitemOpen
  \bibfield  {author} {\bibinfo {author} {\bibfnamefont {M.}~\bibnamefont
  {\ifmmode~\check{S}\else \v{S}\fi{}indik}}, \bibinfo {author} {\bibfnamefont
  {T.}~\bibnamefont {Zawi\ifmmode~\acute{s}\else \'{s}\fi{}lak}}, \bibinfo
  {author} {\bibfnamefont {A.}~\bibnamefont {Recati}},\ and\ \bibinfo {author}
  {\bibfnamefont {S.}~\bibnamefont {Stringari}},\ }\bibfield  {title} {\bibinfo
  {title} {Sound, superfluidity, and layer compressibility in a ring dipolar
  supersolid},\ }\href {https://doi.org/10.1103/PhysRevLett.132.146001}
  {\bibfield  {journal} {\bibinfo  {journal} {Phys. Rev. Lett.}\ }\textbf
  {\bibinfo {volume} {132}},\ \bibinfo {pages} {146001} (\bibinfo {year}
  {2024})}\BibitemShut {NoStop}%
\bibitem [{\citenamefont {Shevchenko}\ and\ \citenamefont
  {Fil}(2007)}]{shevchenko2007}%
  \BibitemOpen
  \bibfield  {author} {\bibinfo {author} {\bibfnamefont {S.~I.}\ \bibnamefont
  {Shevchenko}}\ and\ \bibinfo {author} {\bibfnamefont {D.~V.}\ \bibnamefont
  {Fil}},\ }\bibfield  {title} {\bibinfo {title} {The andreev-bashkin effect in
  a two-component bose gas},\ }\href
  {https://doi.org/10.1134/S106377610707028X} {\bibfield  {journal} {\bibinfo
  {journal} {Journal of Experimental and Theoretical Physics}\ }\textbf
  {\bibinfo {volume} {105}},\ \bibinfo {pages} {135} (\bibinfo {year}
  {2007})}\BibitemShut {NoStop}%
\end{thebibliography}%
\end{document}